\documentclass[aps,preprint]{revtex4}%
\usepackage{amsfonts}
\usepackage{amsmath}
\usepackage{amssymb}
\usepackage{graphicx}%
\begin{document}
\preprint{hep-th/0310192}
\title[Short title for running header]{Mass singularity in QED$_{3}$ }
\author{Yuichi Hoshino}
\affiliation{Kushiro National Colledge of\ technology,Otanoshike nishi 2-32-1,Kushiro 084-0916,Hokkaido,Japan}

\begin{abstract}
Shell document for REV\TeX{} 4.

\end{abstract}
\begin{abstract}
We determine the position space fermion propagator in three dimensional QED
based on Ward-identity and spectral representation.There is a new type of mass
singularity which governs the long distance behaviour.It leads the propagator
vanish at large distancemore strongly than the mass term does.This term
corresponds to Dynamical mass.Momentum space propagator is compared with the
analysis of Schwinger-Dyson equation and our solution contains a
non-perturbative effects beyond the quenched approximation with bare vertex.

\end{abstract}
\volumeyear{year}
\volumenumber{number}
\issuenumber{number}
\eid{identifier}
\date[Date text]{date}
\received[Received text]{date}

\revised[Revised text]{date}

\accepted[Accepted text]{date}

\published[Published text]{date}

\startpage{101}
\endpage{102}
\maketitle
\tableofcontents

\section{Introduction}

In the previous paper we find that the propagator has a new type of mass
singularity in QED$_{3}$[1].The two kinds of self-energy are derived by
Ward-Identity in evaluation of the second order spectral function.To include
intermediate states with infinite numbers of photon we exponentiate the
one-photon matrix element. Usually three dimensional QED is thought as the
high temperature limit of QED in the imaginary time formalism at finite
temperature[2].It is assumed that the three dimensional analysis leads the
leading order of high temperature expansion results in four dimension[3].In
this respect first we study the position space structure of the propagator
especially in the Yennie gauge.In this gauge we can avoid linear infrared
divergences. In three dimension propagator has a logarithmic singularity even
in the Yennie gauge.Fourier transformation to momentum space is done
numerically.Low and high energy behaviour is consistent with other analysis in
Euclidean space[6].In section II Bloch-Nordsieck approximation in three
dimension is reviewed.Section III is for non-perturbative effects.In section
IV mass singularity in four and three dimension is compared and numerical
analysis in momentum space is given.Main results are the followings.

\begin{quotation}
1 Based on Ward-Identity and dispersion-theoretic representation we find
position space propagator $S_{F}(x).$

2 Gauge dependent terms are second-order corrected mass and linear infrared.
divergences which vanishes in the Yennie gauge and others are gauge invariant.

3 Second-order correction to the free propagator yields two kinds of gauge
invariant mass singulairy;position dependent mass and self energy.The
threshold behaviour is determined by self energy and dynamical mass is due to
position dependent mass.Exponentiation of one-photon matrix element we obtain
the full propagator.There exists dynamical symmetry breaking at $e^{2}/8\pi
m=1$ for finite infrared cutoff.
\end{quotation}

\section{Bloch-Nordsieck approximation to $QED_{3}$}

First we consider about the charged praticle which emitt and absorb massless
photons.Usually this process was described by spectral function ;transition
probability of particle into particle and photon state.Multi-photon emitted
from external line is introduced by ladder type diagrams which satisfy
Ward-Identity. Let us beguin by dispersion theoretic description of the
propagator[1]
\begin{align}
S_{F}(p)  &  =\int d^{3}x\exp(ip\cdot x)\left\langle \Omega|T\psi
(x)\overline{\psi}(0)|\Omega\right\rangle \nonumber\\
&  =\int d\omega\frac{\gamma\cdot p\rho_{1}(\omega)+\omega\rho_{2}(\omega
)}{p^{2}-\omega^{2}+i\epsilon},\\
S_{F}(x)  &  =\int d\omega S_{F}(p,\omega).
\end{align}
The field $\psi$ is renormalized and is taken to be a spinor with mass $m$
.Here we introduce intermediate states that contribute the spectral function
\begin{equation}
\sigma(p^{2})=(2\pi)^{2}\sum_{N}\delta^{3}(p-p_{N})\int d^{3}x\exp(ip\cdot
x)\left\langle \Omega|\psi(x)|N\right\rangle \left\langle N|\overline{\psi
}(0)|\Omega\right\rangle .
\end{equation}
Total three-momentum of the state $|N\rangle$ is $p_{N}^{\mu}.$The only
intermediates $N$ contain one spinor and an arbitrary number of
photons.Setting
\begin{equation}
|N\rangle=|r;k_{1},...,k_{n}\rangle,
\end{equation}
where $r$ is the momentum of the spinor $r^{2}=m^{2},$and $k_{i}$ is the
momentum of $i$th soft photon,we have
\begin{align}
\sigma(p^{2})  &  =\int\frac{md^{2}r}{r^{0}}\sum_{n=0}^{\infty}\frac{1}%
{n!}\times(\int\frac{d^{3}k}{(2\pi)^{3}}\theta(k_{0})\delta(k^{2})\sum
_{n})_{n}\delta(p-r-\sum_{i=1}^{n}k_{i})\nonumber\\
&  \times\left\langle \Omega|\psi(x)|r;k_{1},..,k_{n}\right\rangle
\left\langle r;k_{1},..k_{n}|\overline{\psi}(0)|\Omega\right\rangle .
\end{align}
Here the notation
\[
(f(k))_{0}=1,
\]%
\begin{equation}
(f(k))_{n}=\prod_{i=1}^{n}f(k_{i})
\end{equation}
has been introduced.Here we define matrix element
\begin{align}
T_{n}  &  =\left\langle \Omega|\psi|r;k_{1},..k_{n}\right\rangle ,\\
T_{n}^{\mu}  &  =-\int d^{3}x\exp(ik_{n}\cdot x)\left\langle \Omega|T\psi
j^{\mu}|r;k_{1},..k_{n-1}\right\rangle ,
\end{align}
provided
\begin{equation}
\square_{x}T\psi A_{\mu}(x)=T\psi\square_{x}A_{\mu}(x)=T\psi(-j_{\mu
}(x)+\partial_{\mu}^{x}(\partial\cdot A(x))).
\end{equation}
$T_{n}$ satisfies Ward-Identity:
\begin{align}
\partial_{\mu}^{x}T(\psi j_{\mu}(x))  &  =-e\psi(x),\\
\partial_{\mu}^{x}T(\overline{\psi}j_{\mu}(x))  &  =e\overline{\psi
}(x),\nonumber\\
k_{n\mu}T_{n}^{\mu}(r,k_{1},..k_{n})  &  =eT_{n-1}(r,k_{1},..k_{n-1}%
),r^{2}=m^{2}.
\end{align}
Using LSZ the one photon matrix element is given
\begin{align}
T_{1}  &  =\left\langle in|T(\psi_{in}(x),ie\int d^{3}x\overline{\psi}%
_{in}(y)\gamma_{\mu}\psi_{in}(y)A_{in}^{\mu}(y))|r;k\text{ }in\right\rangle
\nonumber\\
&  =ie\int d^{3}yd^{3}zS_{F}(x-y)\gamma_{\mu}\delta^{(3)}(y-z)\exp(i(k\cdot
y+r\cdot z))\epsilon^{\mu}(k,\lambda)U(r,s)\nonumber\\
&  =-ie\frac{(r+k)\cdot\gamma+m}{(r+k)^{2}-m^{2}}\gamma_{\mu}\epsilon^{\mu
}(k,\lambda)\exp(i(k+r)\cdot x)U(r,s),
\end{align}
where $U(r,s)$ is a four-component free particle spinor with positive energy
\begin{equation}
\sum_{S}U(r,s)\overline{U}(r,s)=\frac{\gamma\cdot r+m}{2m}.
\end{equation}
The spectral function $\sigma$ is given by exponentiation of one-photon matrix
element,%
\begin{equation}
T_{n}=%
{\displaystyle\prod\limits_{j=1}^{n}}
T_{1}(k_{j}),
\end{equation}
which yields an infinite laddar approximation
\begin{align}
\sigma(x)  &  =\int\frac{md^{2}r}{r^{0}}\exp(ir\cdot x)\exp(F),\\
F  &  =\int\frac{d^{3}k}{(2\pi)^{2}}\delta(k^{2})\theta(k_{0})\sum_{\lambda
,S}T_{1}(x)T_{1}^{+}(0)\nonumber\\
&  =\int\frac{d^{3}k}{(2\pi)^{2}}\exp(ik\cdot x)\delta(k^{2})\theta
(k_{0})\nonumber\\
&  \times e^{2}tr\left[  \frac{(r+k)\cdot\gamma}{(r+k)^{2}-m^{2}}\gamma^{\mu
}\frac{(r+k)\cdot r}{(r+k)^{2}-m^{2}}\gamma^{\nu}\frac{r\cdot\gamma+m}{2m}%
\Pi_{\mu\nu}\right] \nonumber\\
&  =\int\frac{d^{3}k}{(2\pi)^{2}}\exp(ik\cdot x)\delta(k^{2})\theta
(k_{0})\nonumber\\
&  \times e^{2}\left[  \frac{m^{2}}{(r\cdot k)^{2}}+\frac{1}{r\cdot
k}+(d-1)\frac{\delta(k^{2})}{k^{2}}\right]  .
\end{align}
First we take the trace of $T_{1}T_{1}^{+}$ for simplicity in the infrared.In
this case we take the scalar part of them and assume $\rho_{1}(\omega
)=\rho_{2}(\omega).$In general case there are two kinds of spectral
function.To aviod infared divergences we introduce photon mass $\mu$ as an
infrared cut off.It is helpful to use function $D_{+}(x)$
\begin{align}
D_{+}(x)  &  =\frac{1}{(2\pi)^{2}i}\int\exp(ik\cdot x)d^{3}k\theta
(k^{0})\delta(k^{2}-\mu^{2})\nonumber\\
&  =\frac{1}{(2\pi)^{2}i}\int_{0}^{\infty}J_{0}(kx)\frac{\pi kdk}{2\sqrt
{k^{2}+\mu^{2}}}=\frac{\exp(-\mu x)}{8\pi ix},
\end{align}

to determine $F.$If we use parameter trick
\begin{align}
\lim_{\epsilon\rightarrow0}\int_{0}^{\infty}d\alpha\exp(i(k+i\epsilon
)\cdot(x+\alpha r))  &  =\frac{\exp(ik\cdot x)}{k\cdot r},\\
\lim_{\epsilon\rightarrow0}\int_{0}^{\infty}\alpha d\alpha\exp(i(k+i\epsilon
)\cdot(x+\alpha r))  &  =\frac{\exp(ik\cdot x)}{(k\cdot r)^{2}},
\end{align}
the function $F$ is written in the following form%

\begin{align}
F  &  =ie^{2}m^{2}\int_{0}^{\infty}\alpha d\alpha D_{+}(x+\alpha r,\mu
)-e^{2}\int_{0}^{\infty}d\alpha D_{+}(x+\alpha r,\mu)-ie^{2}(d-1)\frac{
\partial}{\partial\mu}D_{+}(x,\mu)\nonumber\\
&  =\frac{e^{2}m^{2}}{8\pi r^{2}}(-\frac{\exp(-\mu x)}{\mu}+x\operatorname{Ei}
(1,\mu x))+(d-1)\frac{e^{2}}{8\pi\mu}\exp(-\mu x),
\end{align}
where the function $\operatorname{Ei}(n,\mu x)$ is defined
\begin{equation}
\operatorname{Ei} (n,\mu x)=\int_{1}^{\infty}\frac{\exp(-\mu xt)}{t^{n}}dt.
\end{equation}
It is understood that all terms which vanishes with $\mu\rightarrow0$ are
ignored. The leading non trivial contributions to $F$ are
\begin{equation}
\operatorname{Ei} (1,\mu x)=-\gamma-\ln(\mu x)+O(\mu x),
\end{equation}

\begin{align}
F_{1}  &  =\frac{e^{2}m^{2}}{8\pi r^{2}}\left(  -\frac{1}{\mu}+x(1-\ln(\mu
x)-\gamma)\right)  +O(\mu),\nonumber\\
F_{2}  &  =\frac{e^{2}}{8\pi r}(\ln(\mu x)+\gamma)+O(\mu),\nonumber\\
F_{g}  &  =\frac{e^{2}}{8\pi}(\frac{1}{\mu}-x)(d-1)+O(\mu),
\end{align}%
\begin{equation}
F=\frac{e^{2}}{8\pi\mu}(d-2)+\frac{\gamma e^{2}}{8\pi r}+\frac{e^{2}}{8\pi
r}\ln(\mu x)-\frac{e^{2}}{8\pi}x\ln(\mu x)-\frac{e^{2}}{8\pi}x(d-2+\gamma),
\end{equation}
where $\gamma$ is Euler's constant.Using integrals for intermediate state for
on-shell fermion
\begin{align}
\int d^{3}x\exp(ip\cdot x)\int d^{3}r\delta(r^{2}-m^{2})\exp(ir\cdot x)f(r)
&  =f(m),\\
\int\frac{d^{3}x}{(2\pi)^{3}}\exp(ip\cdot x)\int d^{3}r\delta(r^{2}-m^{2})  &
=\frac{1}{m^{2}+p^{2}},
\end{align}
we set $r=m$ in the phase space integral ;
\begin{align}
\sigma(p)  &  =\int\frac{d^{3}x}{(2\pi)^{3}}\exp(ip\cdot x)\int\frac{m}%
{\sqrt{r^{2}+m^{2}}}d^{2}r\exp(ir\cdot x)\exp(F(m,x))\nonumber\\
&  =\int\frac{d^{3}x}{(2\pi)^{3}}\exp(ip\cdot x)\frac{\exp(-mx)}{x}%
\exp(F(m,x)).
\end{align}
Since we evaluated the matrix element $F$ by the on shell limit of fermion and
photon, $F$ can be understood to describes self-energy[5].In section IV and V
we discuss dynamical mass,the renormalization constant ,and bare mass in
connection of each terms. After angular integration we get the propagator%

\begin{equation}
\sigma(p)=\frac{m}{2\pi p}\int_{0}^{\infty}dx\sin(px)\exp(-(m+B)x)(\mu
x)^{-Cx+D},
\end{equation}
where
\begin{equation}
A=\frac{e^{2}}{8\pi\mu}(d-2)+\frac{\gamma e^{2}}{8\pi m},B=\frac{e^{2}}{8\pi
}(d-2+\gamma),C=\frac{e^{2}}{8\pi},D=\frac{e^{2}}{8\pi m}.
\end{equation}
Here we show up to $O(e^{2})$ propagator $\sigma(p)$
\begin{align}
\sigma^{(2)}(p)  &  =\int\frac{d^{3}x}{(2\pi)^{3}}\exp(ip\cdot x)\int
\frac{md^{2}r}{r^{0}}\exp(ir\cdot x)F(x)\nonumber\\
&  =\frac{m}{2\pi p}\int_{0}^{\infty}dx\sin(px)\exp(-mx)[A-Bx-Cx\ln(\mu
x)+D\ln(\mu x)]\nonumber\\
&  =[\frac{2m(1+A)}{m^{2}+p^{2}}-\frac{m^{2}B}{(m^{2}+p^{2})^{2}}%
+2m(DI_{1}-CI_{2})],
\end{align}
where $I_{1},I_{2}$ are the following integrals
\begin{align}
I_{1}  &  =\int_{0}^{\infty}\frac{\sin(px)\exp(-mx)}{p}\ln(\mu x)dx\nonumber\\
&  =\frac{-\gamma}{m^{2}+p^{2}}-\frac{\ln((m^{2}+p^{2})/\mu^{2})}%
{2(m^{2}+p^{2})}-\frac{\ln((m-\sqrt{-p^{2}})/(m+\sqrt{-p^{2}}))}{m^{2}+p^{2}%
},\\
I_{2}  &  =\int_{0}^{\infty}\frac{\sin(px)\exp(-mx)}{p}x\ln(\mu
x)dx\nonumber\\
&  =\frac{-m}{(m^{2}+p^{2})^{2}}[\ln((m-\sqrt{-p^{2}})/(m+\sqrt{-p^{2}}%
))+\ln((m^{2}+p^{2})/\mu^{2})-2(1-\gamma)].
\end{align}
Notice that the Yennie gauge($d=2)$ is free from linear infrared divergences
but associates a logarithmic divergences.In four dimension logarithmic
divergence disappear and we have a free propagator in this gauge.This point is
a clear difference between three and four dimensional case. For definitteness
hereafter we choose the Yennie gauge.In this gauge the position space
propagator is
\begin{equation}
S_{F}(x)=-\int(i\gamma\cdot\partial+\omega)\frac{\exp(-\omega x)}{4\pi
x}\sigma(\omega)d\omega,
\end{equation}
and it is written in momentum space
\begin{align}
S_{F}(p)  &  =\int d\omega\frac{(\gamma\cdot p+\omega)\sigma(\omega)}%
{p^{2}-\omega^{2}+i\epsilon}=\frac{\gamma\cdot p}{p}\sigma(p)+\sigma
(p),\nonumber\\
\sigma(p)  &  =\int d^{3}x\exp(ip\cdot x)\sigma(x),\\
\sigma(x)  &  =\frac{m\exp(-(m+B)x}{4\pi x}(\mu x)^{-Cx+D}.
\end{align}
Finally we show the general spin dependent spectral function in the $O(e^{2})$
for $d=2$ gauge%
\begin{equation}
F=(\gamma\cdot r+m)[\frac{e^{2}}{8\pi}(-\gamma x-x\ln(mx)+\frac{e^{2}}{8\pi
m}(\ln(mx)+\gamma)]-\gamma\cdot\gamma\frac{e^{2}}{8\pi m^{2}x}.
\end{equation}
The first term is the same as the preavious one.The second term is independent
of the cutoff $\mu$ and not significant in the infrared.In Euclidean space if
we exponentiate it that becomes%
\begin{align}
&  \int\frac{mr^{2}dr}{\sqrt{r^{2}+m^{2}}}\exp(ir\cdot x)\frac{\exp(-mx)}{2\pi
x}(1+\frac{ir\cdot\gamma}{\sqrt{-r^{2}}})\exp((\frac{ir\cdot\gamma}%
{\sqrt{-r^{2}}}+1)F)\nonumber\\
&  =\int\frac{mr^{2}dr}{\sqrt{r^{2}+m^{2}}}\exp(ir\cdot x)\frac{\exp
(-mx)}{2\pi x}(1+\frac{ir\cdot x}{\sqrt{-r^{2}}})\exp(\frac{F}{2})(\cosh
(\frac{F}{2})+i\frac{\sigma\cdot r}{\sqrt{-r^{2}}}\sinh(\frac{F}{2})).
\end{align}
This form corresponds to the infinite ladder graph of the fermion propagator
with renormalized mass $m$ and corrected vertex$.$In section IV we discuss
physical meanings of each terms in $\sigma(x)$ and the dynamical effects in
momentum space numerically.

\section{Vanishing bare mass and vacuum expectation value of condensate}

In this section we examine the renormalization constant and bare mass and
study the condition of vanishing bare mass based on spectral
representation.The spinor propagator in position space is expressed in the
fllowing[3]
\begin{align}
S_{F}(x)  &  =-\int(i\gamma\cdot\partial+\omega)\frac{\exp(-\omega x)}{4\pi
x}\sigma(\omega)d\omega\\
S_{F}(p)  &  =\frac{\gamma\cdot p}{p}\sigma(p)+\sigma(p).
\end{align}
First term is related to wave function renormalization and second term is a
mass function which contains renormalized and dynamical mass.The equation for
the renormalization constant in terms of the spectral functions read%

\begin{equation}
\lim_{p\rightarrow\infty}\frac{Z_{2}^{-1}(\gamma\cdot p+m_{0})}{p^{2}%
-m^{2}+i\epsilon}=\lim_{p\rightarrow0}\int\frac{(\gamma\cdot p+\omega
)\sigma(\omega)d\omega}{p^{2}-\omega^{2}+i\epsilon}.
\end{equation}
Instead we determine them directly by taking the high energy limit of $S_{F}$%
\begin{align}
m_{0}Z_{2}^{-1}  &  =\lim_{p\rightarrow\infty}p^{2}\frac{1}{4}tr[S_{F}(p)],\\
Z_{2}^{-1}  &  =\lim_{p\rightarrow\infty}\frac{1}{4}tr[\gamma\cdot pS_{F}(p)].
\end{align}
To determine $Z_{2}^{-1},$first we show free case
\begin{equation}
S_{F}^{(0)}(x)=(i\gamma\cdot\partial)\frac{1}{4\pi\sqrt{-x^{2}}}=\frac
{i\gamma\cdot x}{4\pi(-x^{2})^{3/2}}=-\frac{i\gamma\cdot x}{\sqrt{-x^{2}}%
}\frac{1}{4\pi x^{2}},
\end{equation}
where the dimension of the $S_{F}^{(0)}(x)$ is equal to $1/x^{2}.$ In momentum
space
\begin{equation}
\frac{1}{p}\int_{0}^{\infty}\frac{x\sin(px)}{p}x^{D-1}dx\sim p^{-3-D},
\end{equation}
and this shows the ordinary expression
\begin{equation}
\lim_{p\rightarrow\infty}\frac{1}{4}tr(\gamma\cdot pS_{F}(p))\sim p^{-1-D}.
\end{equation}
In this way we obtain
\begin{equation}
Z_{2}^{-1}\sim\lim_{p\rightarrow\infty}p^{-1-D}=\left(
\begin{array}
[c]{cc}%
0 & (0<D)\\
1 & (0=D)
\end{array}
\right)  .
\end{equation}
In the same way to evaluate $Z_{2}^{-1},$bare mass reads
\begin{align}
\lim_{p\rightarrow\infty}\frac{1}{4}p^{2}tr(S_{F}(p))  &  =m\lim
_{p\rightarrow\infty}p^{2}\int_{0}^{\infty}\frac{x\sin(px)}{px}x^{D-1}%
dx\nonumber\\
&  \sim mp^{-D},
\end{align}
in the high energy limit if we use the formula%
\begin{equation}
m_{0}Z_{2}^{-1}\sim m\lim_{p\rightarrow\infty}p^{-D}=0\text{ },
\end{equation}
and we obtain
\begin{equation}
\frac{m_{0}Z_{2}^{-1}}{m}/Z_{2}^{-1}\geq0.
\end{equation}
Usually mass is a parameter which appears in the Lagrangean.For example chiral
symmetry is defined for the bare quantity.In ref[4] the relation between bare
mass and renormalized mass in the Schwinger-Dyson equation is discussed based
on renormalization group equation and shown that the bare mass vanishes in the
high energy limit even if we start from the finite bare mass in the theory.It
suggests that symmetry properties must be discussed in terms of renormalized
quanties.In QCD bare mass vanishes in the short distance by asymptotic
freedom.And Dynamical mass vanishes too.In our apprximation this problem is
understood that the two kinds of mass are always generated in which the second
order terms in the coupling constans is gauge dependent but another is gauge
invariant.In our position space propagator we can easily see the differences
between bare mass and dynamical mass.But in momentum space its relation is
complicated.In this case the symmetry of massless fermion is realized in the
short distance. Condition of the vanishment of the bare mass depends on the
ratio of the renormalized mass and fixed coupling constant.There is a chiral
symmetry at short distance where the bare or dynamical mass vanishes but its
breaking must be discussed in terms of the values of the order parameter.
Therefore it is interesting to study the possibility of pair condensation in
our approximation.The vacuum expectation value of pair condensate is
evaluated
\begin{align}
\left\langle \overline{\psi}\psi\right\rangle  &  =-trS_{F}(x)=-2m\lim
_{x\rightarrow0_{+}}\frac{\exp(-mx)}{x}(\mu x)^{-Cx+D}\nonumber\\
&  =\left(
\begin{array}
[c]{ll}%
0 & (1<D)\\
finite & (1=D)\\
\infty & (1>D)
\end{array}
\right)  ,
\end{align}
for fixed $\mu.$We do not see confinement($Z_{2}\neq0)$ but
condentation(finite $\left\langle \overline{\psi}\psi\right\rangle )$ occurs
at $D=1.$It is understood that in QCD the anomalous dimension of quark or $D$
in our model vanishes in the high energy limit thus $Z_{2}^{-1}=$finite.In
that case the vanishment of the bare mass is automatically satisfied[4].In the
weak coupling limit we obtain $Z_{2}=1,m_{0}=m$ and $\left\langle
\overline{\psi}\psi\right\rangle =\infty$ . If we introduce chiral symmetry as
global $SU(n)_{L}\times SU(n)_{R}$ ,it breaks dynamically into $SU(n)_{V}$ as
in QCD[3] for $D=1$.Our model is also applicable to relativistic model of
super fluidity in three dimension.

\section{Mass singularity in momentum space}

In this section we study the effect of position depedent mass,self energy in
momentum space%
\begin{align}
M(x)  &  =\frac{e^{2}}{8\pi}\ln(\mu x),\\
self\text{ }energy  &  =\frac{e^{2}}{8\pi}\ln(\mu x).
\end{align}
The position space free propagator
\begin{equation}
S_{F}(x,m_{0})=-(i\gamma\cdot\partial+m_{0})\frac{\exp(-m_{0}x)}{4\pi x}%
\end{equation}
is modified by these two terms which are related to dynamical mass and wave
function renormalization.To see this let us think about the effect of spectral
function in position space propagator%
\begin{equation}
\sigma(x)=\frac{\exp(-(m_{0}+\frac{e^{2}\gamma}{8\pi})x)}{4\pi x}(\mu
x)^{-\frac{e^{2}}{8\pi}x}(\mu x)^{\frac{e^{2}}{8\pi m}}.
\end{equation}
It is easy to see that the probability of particles which are separated with
each other in the long distance is supressed by the factor $(\mu x)^{-Cx},$and
the self energy modifies the short distance behaviour from the bare $1/x$ to
$1/x^{1-D}$.The effect of self energy for the infrared behaviour of the free
particle with mass $m$ can be seen by its fourier transform[1]
\begin{align}
\int_{0}^{\infty}x^{2}\frac{\sin(px)}{px}\frac{\exp(-mx)}{4\pi x}(\mu
x)^{a}dx  &  =-\Gamma(a+1)\cos(\frac{\pi a}{2})(p^{2}+m^{2})^{-1-a/2}\mu
^{a}\nonumber\\
&  \times\lbrack(\sqrt{-p^{2}}+m)(\frac{\sqrt{-p^{2}}-m}{\sqrt{-p^{2}}%
+m})^{-a/2}+(\sqrt{-p^{2}}-m)(\frac{\sqrt{-p^{2}}+m}{\sqrt{-p^{2}}-m}%
)^{-a/2}]\nonumber\\
&  \sim(\sqrt{-p^{2}}-m)^{-1-a}\text{ near }p^{2}=m^{2}.\text{ }%
\end{align}
Usually constant $D$ represents the coefficent of the leading infrared
divergence for fixed mass in four dimension. Therefore self energy has the
same effects in three dimension as in four dimension[5].In the non-linear
Dyson-Schwinger equationof the fermion propagator or operator product
expansion we can evaluate the dynamical mass which depends on momentum[4].Here
we show that $M(x)$ satisfies the same property as the dynamical mass.Let us
define fourier transform of the scalar part of the propagator
\begin{equation}
F.T(\frac{\exp(-mx)}{4\pi x}(\mu x)^{-Cx+D})=\sigma_{E}(p)=\frac{m}{2\pi
p}\int\sin(px)\exp(-mx)(\mu x)^{-Cx+D}dx.
\end{equation}
The momentum dependence of $\sigma_{E}(p)$ agrees well to the numerical
solution of the Dyson-Schwinger equation with vanishing bare mass in the
Landau gauge.It dumps as $p^{-2-D}$ and stays constant at small $p[6].$It is
easy to see that short distance behaviour is descrived by wave renormalization
part $(\mu x)^{D\text{ }}$and the long distance behaviour is given by
$\exp(-M(x)x)=(\mu x)^{-Cx}$.The effect of $(\mu x)^{-Cx}$, first we check the
$o(e^{2})$ contribution%
\begin{align}
\sigma_{M}(p)  &  =-\frac{e^{2}}{8\pi}I_{2},\\
I_{2}  &  =\int_{0}^{\infty}x^{2}\frac{\sin(px)}{px}\frac{\exp(-mx)}{4\pi
x}x\ln(\mu x)dx\nonumber\\
&  =\frac{-m}{(p^{2}+m^{2})^{2}}[\ln(\frac{(m-\sqrt{-p^{2}})}{(m+\sqrt{-p^{2}%
})})+\ln(\frac{(m^{2}+p^{2})}{\mu^{2}})-2(1-\gamma)].
\end{align}
Thus a secon order correction of the propagator dumps as $(O(1)+\ln(p^{2}%
/\mu^{2}))me^{2}/p^{4}$at high energy. It is not easy to see the effect of the
factor $(\mu x)^{-Cx}$in momentum space.Therefore the scalar part of the
propagator $m/(m^{2}+p^{2})$ receives an additional correction by $(\mu
x)^{-Cx}.$If the dynamical symmetry breaking and mass generation occur it has
been discussed that the propagator will have a branch point on the real
$p^{2}$ axis[7].However the analysis in Minkowski space is difficult due to
the fact that the function $(\mu x)^{-Cx}$ cannot be integrated
analytically.Therefore we see only the perturbative resutls.For this purpose
it may be helpful to study the minkowski space by replacement from Euclidean
distance $R=\sqrt{x^{2}}$ $\rightarrow$ $iT$ in $\sigma(x)[8]$,%
\begin{align}
\sigma(T)  &  \simeq\frac{\exp(-imT)}{2\pi m}(\mu iT)^{-iCT+D}=\frac
{\exp(-imT)}{2\pi m}(\mu T)^{-iCT+D}\exp(\frac{\pi}{2}i(-iCT+D))\\
&  =\frac{\exp(\frac{\pi}{2}(CT+Di))(\mu T)^{D}}{4\pi}(\cos(CT\ln(\mu
T)+mT)-i\sin(CT\ln(\mu T)+mT),
\end{align}
provided%
\begin{equation}
(i\mu T)^{-iCT+D}=\exp(\frac{\pi}{2}(C+Di))(\mu T)^{D}(\cos(CT\ln(\mu
T))-i\sin(CT\ln(\mu T))).
\end{equation}
Here we notice that $\sigma(T)$ becomes complex valued function.Fourier
transform into energy is given%
\begin{align}
\sigma(E)  &  =\lim_{T\rightarrow\infty}\frac{1}{2T}\int_{-T}^{T}%
\cos(ET)\sigma(T)dT,\\
\int dE\sigma(E)  &  =-\lim_{T\rightarrow\infty}\frac{1}{2T}\int_{-T}^{T}%
\frac{\sin(ET)}{T}\sigma(T)dT\nonumber\\
&  =-\lim_{T\rightarrow\infty}\frac{1}{2T}\pi\int_{-T}^{T}\delta
(E)\sigma(T)dT.
\end{align}
In free case integration over space gives
\begin{equation}
\sigma_{0}(t)=\int dx^{2}\frac{\exp(-m\sqrt{-t^{2}+x^{2}})}{4\pi\sqrt
{-t^{2}+x^{2}}}=\frac{\exp(-imt)}{2\pi m}.
\end{equation}%
\[
\int\sigma_{0}(E)dE=\lim_{T\rightarrow\infty}\frac{1}{2T}\frac{2\pi\sin
(mT)}{\pi m^{2}}%
\]
Thus energy distribution for free fermion becomes%
\begin{align}
\sigma_{0}(E)  &  =\frac{1}{2T}\int_{-T}^{T}\cos(Et)\sigma_{0}(t)dt\nonumber\\
&  =\frac{m\sin(mT)\cos(ET)+E\cos(mT)\sin(ET)}{2\pi Tm(E^{2}-m^{2}+i\epsilon
)}.
\end{align}
Taking imaginary part%
\begin{equation}
\Im\sigma_{0}(E)=\frac{\sin(2mT)}{2T}%
\end{equation}
gives the residue at $E^{2}=m^{2}.$In free case imaginary part is
$\delta(E^{2}-m^{2})$ and the continium contribution is given by an integral
of the principal part in the region given by $\theta(E^{2}-m^{2})$.By
numerical analysis using Maple we find that $\sigma(E)$ osscillates and
diverges in the region
\begin{equation}
E\simeq m_{eff}=m_{0}+\frac{e^{2}\gamma}{8\pi}+C
\end{equation}

\section{Summary}

Infrared behaviour of the propagator was examined in the Bolch-Nordsieck
approximation to QED$_{3}$. Even in the Yennie gauge there is a logarithmic
infrared divergence but free from linear infrared divergences.Position
dependent mass and self energy are both gauge independent in our
approximation.Therefore in three dimension we examine the structure of the
propagator in this gauge.In our approximation confinement property is given by
position dependent mass.We evaluated the renormalization constant and bare
mass based on spectral function sum rule and dynamical mass in momentum space
is analised with finite infrared cutoff. In our approximation there seems to
be a critical coupling constant for vacuum expectation value $\left\langle
\overline{\psi}\psi\right\rangle $ which is due to wave function
renormalization.This is a clear difference between lowest laddar approximation
to the Schwnger-Dyson equation with bare vertex and ours.In Euclidean space we
see the momentum dependence of the propagator which has been known in the
anlysis of the Schwinger-Dyson equations except for the wave function renormalization[6].

\section{References}

\noindent\lbrack1]Y.Hoshino$,$JHEP05(2003)
075;R.Jackiw,L.Soloviev,Phys.Rev.\textbf{173}.(1968)1458.\newline%
[2]S.Deser,R,Jackiw,S.Templton,Ann.Phys.(NY)\textbf{140}(1982)372.\newline%
[3]T.Appelquist,R.Pisarski,Phys.Rev.\textbf{D23}%
(1981)2305;T.Appelquist,U.Heinz,Phys.Rev.$\mathbf{D24}$ .(1981)2305.\newline%
[4]K.Nishijima,Prog.Theor.Phys.\textbf{81}
(1989)878;K.Nishijima,Prog.Theor.Phys.\textbf{83}(1990)1200.\newline%
[5]L.S.Brown,Quantum Field Theory,Cambridege University Press(1992).\newline%
[6]T.Appelquist,D.Nash,L.C.R.Wijewardhana,Critical behaviour in
(2+1)-dimensional QED,Phys.Rev.Lett.\textbf{60}%
(1988)1338;Y.Hoshino,T.Matsuyama,Phys.Lett.\textbf{B222}(1989)493.\newline%
[7]D.Atkinson,D.W.E.Blatt,Nucl.Phys.\textbf{151B}%
(1979)342;P.Maris,Phys.Rev.\textbf{D52}(1995);

Y.Hoshino,Il.Nouvo.Cim\textbf{.112A}(1999)335.\newline%
[8]J.Shwinger,Particles,Sources,and Fields,volume I(1970),Perseus Books Publishing,L,L,C.

\end{document}